\documentclass[12pt,preprint]{aastex}

\begin{document}

\def\K{{\rm\,K}}
\def\cm{{\rm\,cm}}
\def\sec{{\rm\,s}}
\def\ergs{{\rm\,ergs}}

\title{Compton Heated Outflow from CDAFs}

\author{Myeong-Gu Park}
\affil{Department of Astronomy and Atmospheric Sciences,
                 Kyungpook National University, Daegu 702-701, KOREA}
\email{mgp@knu.ac.kr}
\and
\author{Jeremiah P. Ostriker}
\affil{Princeton University Observatory, Princeton University,
       Princeton, NJ 08544; \\
       Institute of Astronomy, Cambridge, UK}
\email{jpo@astro.princeton.edu}

\shorttitle{Wind in CDAF}
\shortauthors{Park \& Ostriker}

\begin{abstract}
Convection-dominated accretion flows (CDAF) are expected to have a
shallower density profile and a higher radiation efficiency as compared
to advection-dominated accretion flows (ADAF). Both solutions have been
developed to account for the observed properties of the low luminosity,
high temperature X-ray sources believed to involve accretion onto
massive black holes. Self-similar CDAFs also have steeper poloidal
density gradients and temperatures close to the virial temperature. All
these characteristics make CDAFs more capable of producing polar
outflows driven by Compton heating as compared to other classical
accretion disks. We investigate the conditions for producing such
outflows in CDAFs and look for the mass accretion rate, or, equally,
the luminosity of CDAFs for which such outflows will exist. When the
electron temperature saturates around $10^{11}\K$ at the inner region,
polar outflows are probable for $8\times10^{-7} \lesssim L/L_E \lesssim
4\times10^{-5}$, where \(L_E\) is the Eddington luminosity. Outflows
are well collimated with small opening angles. The luminosity range for
which outflow solutions exist is narrower for lower electron
temperature flows and disappears completely for electron temperature
$\lesssim 6\times 10^9\K$. When the magnetic field is present, we find
that outflows are possible if the magnetic field is less than from 10\%
to 1\% of the equipartition field. We also find that outflows are more
likely to be produced when the viscosity parameter \(\alpha\) is small.
The tendency for jet-like collimated outflows for these solutions is
presumably astrophysically relevant given the high frequency of jets
from AGNs.
\end{abstract}


\section{Introduction}

Advection-dominated accretion flows (ADAF) nicely complement the
classic thin disk accretion flows (Shakura \& Sunyaev 1973; Ichimaru
1977; Rees et al. 1982; Narayan \& Yi 1994, 1995; Abramowicz et al.
1995), and have been successfully applied to variety of cosmic objects,
from galactic X-ray binaries to diffuse X-ray background (see Narayan,
Mahadevan, \& Quataert 1999 for review). However, analytic studies of
ADAFs indicated the convectively unstable nature of ADAFs (Narayan \&
Yi 1994, 1995a; see Begelman \& Meier 1982 for radiation-dominated
ADAF), which has subsequently been proved in a series of numerical
simulations (Igumenshchev, Chen, \& Abramowicz 1996; Igumenshchev \&
Abramowicz 1999, 2000; Stone, Pringle, \& Begelman 1999; Igumenshchev,
Abramowicz, \& Narayan 2000). Especially, the numerical studies of
ADAFs by Igumenshchev \& Abramowicz (1999, 2000) show that an ADAF
becomes a convection-dominated accretion flow (CDAF) whenever the
viscosity parameter $\alpha \lesssim 0.1$. Further analyses of
self-similar CDAF solutions show their unique properties (Narayan,
Igumenshchev \& Abramowicz 2000, hereafter NIA; Quataert \& Gruzinov
2000, hereafter QG): the density varies as $\rho \propto R^{-1/2}$ ($R$
is the radius), much flatter than the usual $R^{-3/2}$ in an ADAF or in
spherical accretion. Correspondingly, the mean radial velocity varies
as $v \propto R^{-3/2}$, compared to $R^{-1/2}$ in ADAF or in spherical
flow. Energy generated at the inner part of the flow is transported to
the outer part by convection. A CDAF has perhaps as much resemblance to
the rotating stellar envelope of a massive star as to the usual
accretion flow.

The other aspect of the two-dimensional nature of ADAF or CDAF
solutions, often neglected, is the interaction between the outgoing
radiation produced at smaller radii with the inflowing gas in the outer
part of the flow. In optically thick stars this interaction plays a
vital role in establishing the equilibrium states. The radiative
interaction also plays a very important role in pure spherical
accretion flows (Ostriker et al. 1976; Cowie, Ostriker, \& Stark 1978;
Wandel, Yahil, \& Milgrom 1984; Park 1990a, 1990b; Nobili, Turolla, \&
Zampieri 1991; Zampieri, Miller, \& Turolla 1996; Ciotti \& Ostriker
1997, 2001). Park \& Ostriker (1999, 2001) studied the same radiative
interaction in the context of the ADAF solution and found that a polar
outflow can be generated through Compton heating of electrons by
high-energy photons emitted by the inner, hot part of the flow. The
winds generated by the processes in the papers listed above are not
momentum driven. Rather, they are caused by overheating of the gas in
the slowly moving, low density polar regions. Outflows may also be
generated from ADAF by other hydrodynamic processes (Narayan \& Yi
1995; Xu \& Chen 1997; Blandford \& Begelman 1999).

In this work, we study the conditions for CDAFs to develop radiation
driven outflows. We adopt the self-similar CDAF solution as the
background flow structure (NIA; QG). The treatment in this paper is two
dimensional, adopting the angular profile of NIA and QG, except that
the radiation field is simplified and treated as spherically symmetric.

\section{Flow Properties}

\subsection{Density and Temperature}

Multi-dimensional numerical simulations and analytic analyses show that
density and temperature profiles of CDAF follow a self-similar form in
radius (NIA; QG; Ball, Narayan \& Quataert 2001, hereafter BNQ):
\begin{eqnarray}\label{eq:rho_Ti}
      \rho(r) &\propto& r^{-1/2} \\
      T_i(r) &\propto& r^{-1},
\end{eqnarray}
where $r$ is the radius in units of the Schwarzschild radius, $R_S
\equiv 2GM/c^2 = 3.0\times10^5 m \cm$, and $m$ is the black hole mass
in solar units. This shallow radial dependence corresponds to marginal
stability to convection. QG further showed that the density should
follow a power law in $\sin \vartheta$ ($\vartheta$ is the angle from
the pole) while the temperature is constant on spherical shells. So we
shall assume that the density is given by
\begin{eqnarray}\label{eq:rho_r_th}
      \rho (r,\vartheta) &=& \rho_0 r^{-1/2} \sin ^2 \vartheta \\
      &=& \rho_{out} ( r / r_{out} )^{-1/2} \sin ^2 \vartheta.
\end{eqnarray}
The flow extends from $r_{out}$ down to $r_{in}=1$, with the outer
radius (analogous to the stellar photosphere) fixed by energy balance
considerations.

The electron temperature $T_e$ in an accretion flow is determined by
the balance between various cooling and heating processes. In general,
detailed studies of hot accretion flows (Narayan \& Yi 1995b; Narayan,
Barret, \& McClintock 1997; BNQ) show that the electron temperature is
equal to the ion temperature for $T_e \lesssim 10^9 \K$, and then, at
smaller radii, due to highly efficient relativistic bremsstrahlung and
synchrotron emission, it flattens to somewhere between $10^9\K$ to
$10^{11}\K$, depending on the amount of direct viscous heating to
electrons. Hence, in this paper, we approximate the electron
temperature as
\begin{eqnarray}\label{eq:Te_r}
      T_e (r,\vartheta) &=& \frac{T_0}{r} \quad\hbox{for}\quad r > r_1 \\
                       &=& T_1 \quad\hbox{for}\quad r \leq r_1 ,
\end{eqnarray}
where $T_0 \approx 10^{12}\K$, $T_1$ is some constant between $10^9 \K$
and $10^{11} \K$, and $r_1 \equiv T_0/T_1$.

From equation (\ref{eq:rho_r_th}), we find that the total Thomson
optical depth along a given direction \(\vartheta\) is
\begin{equation}\label{eq:taues}
   \tau_{es} \simeq \frac{1}{2}\dot m r_{out}^{1/2} \sin^2\vartheta,
\end{equation}
where $\dot m \equiv \dot M / \dot{M_E} = \dot M c^2 /L_E$ is the
dimensionless mass accretion rate. Our analysis, which is to be applied
to low accretion rate solutions, will assume that the flow is optically
thin to both scattering and redistribution processes.

\subsection{Radiation Field and Radiation Temperature}

The inner part of the accretion flow generally produces higher energy
photons, while the outer part produces lower energy photons. Hence the
radiation field at a given position is determined by the contribution
from inner radiating shells plus the outer shells. Although in each
shell, the density, and therefore, the emissivity (i.e. cooling
function) is a function of poloidal angle $\vartheta$, the energy
density is more uniform over angle than is the emissivity. So, we will
simplify the calculation of radiative transfer by assuming each
radiating shell is homogeneous, using an angle average over the sphere.

Therefore, the radiation energy density at given $r$, $E_X(r)$, is given by
\begin{equation}\label{eq:EXr}
c E_X(r) = R_s \left[
           \frac{1}{4\pi r^2}\int_{r_{in}}^r \Lambda(r')4\pi r'^2 dr'
         + \int_r^{r_{out}} \Lambda(r')
           \frac{r'}{r}\ln\sqrt{\frac{r'+r}{r'-r}} dr' \right],
\end{equation}
where the first term is the contribution from inner radiating shells
and the second term from outer radiating shells (see Appendix A). The
emissivity per unit volume is denoted as the cooling function
$\Lambda$, which will be described in \S3.1.

For relativistic un-Comptonized bremsstrahlung, with $\Lambda \propto
\rho^2 T_e$ and CDAF scaling $\rho \propto r^{-1/2}$ and $T_e \propto
r^{-1}$, the incremental luminosity $dL_X(r)/dr = \hbox{constant}$, and
the luminosity at a given radius $r$ has the largest contribution from
the outermost shell. Nevertheless, equation (\ref{eq:EXr}) gives a peak
energy density in the inner region (cf. {\sl dashed line} in Fig. 1)
and a significant radiation pressure gradient. In addition, when there
is strong Comptonization, the contribution from the inner region can be
significantly enhanced.

Moreover, the amount of Compton heating is determined by the luminosity
times the photon energy, and is largely dominated by the inner hot
region. The radiation temperature, defined as $kT_X(r)$ being the
energy-weighted mean photon energy, at a given radius $r$ is determined
by the equation
\begin{equation}\label{eq:TXr}
c T_X(r) E_X(r) = R_S \left[
                  \frac{1}{4\pi r^2}\int_{r_{in}}^r T_X(r')\Lambda(r')
                  4\pi r'^2 dr' + \int_r^{r_{out}} T_X(r')\Lambda(r')
                  \frac{r'}{r}\ln\sqrt{\frac{r'+r}{r'-r}} dr' \right].
\end{equation}
This quantity is more concentrated than $E_X(r)$, especially in the
presence of strong Comptonization [see {\sl dotted curve} in Fig. 1 for
$T_X(r)$].

In presence of an overheated wind, some region along the pole in CDAF
will not be able to accrete matter, while the remaining equatorial
region normally accretes (see \S3.3). In most of the relevant parameter
space, the overheated `funnel' is very narrow, and we assume (with
regard to energy generation) that the self-similar CDAF flow is filling
the whole space including the polar region, omitting the small
correction due to the solid angle $\Omega_W$ ($\sim 0.2$ sr) occupied
by the outflowing wind/jet.

The resulting profiles of radiation moments and the radiation
temperature show the aforementioned characteristics of CDAFs. The
luminosity,
\begin{equation}\label{eq:Lr}
  L(r) = R_S^3 \int_{r_{in}}^r \Lambda(r')4\pi r'^2 dr',
\end{equation}
increases at large radius ({\sl solid curve} in Fig. 1), while the
radiation temperature is roughly flat or decreasing slowly outward
after peaking at some intermediate radius ({\sl dotted curve} in Fig.
1).

\subsection{Outer Boundary}

The mathematical self-similar CDAF has zero mass accretion rate (NIA),
and this, of course, corresponds to zero luminosity. However, the
accretion flow must lose energy by radiative cooling, and this energy
loss is very likely provided by the convective energy transport (NIA,
BNQ). We follow BNQ to assume that the total luminosity of ADAF is
proportional to the mass accretion rate:
\begin{equation}\label{eq:LvsMdot}
   L(r_{out}) = \eta_c \epsilon_c \dot{M} c^2 ,
\end{equation}
where the convective efficiency $\epsilon_c$, determined by numerical
simulations to be in the range $\sim 10^{-2} - 10^{-3}$ and $\eta_c$ is
the fraction of convected energy to be radiated away by electrons. In
subsequent examples, we nominally adopt $\epsilon_c \eta_c = 10^{-2}$
(BNQ). We also tried smaller values of $\epsilon_c \eta_c$ or a scaling
of $\epsilon_c \eta_c$ with $\dot m$ (cf. \S3.3), and the results were
qualitatively the same.

We determine the outer boundary for a given mass accretion rate $\dot
m$ by finding the radius $r_{out}$ at which equation (\ref{eq:LvsMdot})
is satisfied. This gives a somewhat different relation between the
total luminosity and the mass accretion rate from that in BNQ because
we have used Comptonized relativistic bremsstrahlung (and synchrotron)
while BNQ employ only un-Comptonized non-relativistic bremsstrahlung.
However, the general characteristic of the relation is the same: higher
mass accretion flows have smaller outer boundary radii. This is due to
the special radial profile of density in CDAF solutions. Because of the
weak density gradient with radius, most of the luminosity contribution
is from the outer part of the flow, which makes CDAF flows quite
distinct from other accretion flows including ADAFs. The dimensionless
total luminosity, $L (r_{out}) / L_E$, is directly proportional to the
dimensionless mass accretion rate $\dot m \equiv \dot M / \dot M_E$ by
equation (\ref{eq:LvsMdot}) while the luminosity accumulated from
bremsstrahlung is very roughly proportional to $\dot m^2 r_{out}$ (with
a higher power in $\dot m$ if Comptonization is important). Since these
two luminosities must be equal, a higher $\dot m$ flow implies a
smaller $r_{out}$ (see {\sl solid lines} in Fig. 2). Therefore lower
mass accretion rate CDAF solutions have lower luminosity, yet are more
extended, if we treat $\epsilon_c \eta_c$ as fixed, independent of
$\dot m$.

\section{Outflow}

\subsection{Cooling and Heating}

The main cooling mechanism we consider in this work is Comptonized
bremsstrahlung and Comptonized synchrotron. We first focus on the
Comptonized bremsstrahlung. This process is not only the source of
electron cooling but also the source of the Compton heating radiation
field.

The Comptonized bremsstrahlung cooling rate per unit volume can be
reasonably approximated by (Svensson 1982; Stepney \& Guilbert 1983)
\begin{equation}\label{eq:lamb_br}
   \Lambda_{br} = \sigma_T c \alpha_f m_e c^2 n^2 [ F_{ei}(T_e) +
   F_{ee}(T_e) ],
\end{equation}
where
\begin{eqnarray}\label{eq:FeiFee}
   F_{ei} & = & 4 \left(\frac{2}{\pi^3}\right)^{1/2}
                \theta_e^{1/2}(1+1.781\theta_e^{1.34})
          \quad\hbox{for}~~ \theta_e < 1 \\
          & = & \frac{9}{2\pi} \theta_e
                \left[\ln(1.123\theta_e+0.48)+1.5\right]
          \quad\hbox{for}~~ \theta_e > 1 \nonumber\\
   F_{ee} & = & \frac{5}{6\pi^{3/2}}(44-3\pi^2)\theta_e^{3/2}
                (1+1.1\theta_e+\theta_e^2-1.25\theta_e^{5/2})
          \quad\hbox{for}~~ \theta_e < 1 \\
          & = & \frac{9}{\pi} \theta_e
                \left[\ln(1.123\theta_e)+1.2746\right]
          \quad\hbox{for}~~ \theta_e > 1 ,\nonumber
\end{eqnarray}
$\theta_e \equiv k T_e / m_e c^2$, $n$ is the electron (ion) number
density, $\sigma_T$ the Thomson cross section, $\alpha_f$ the fine
structure constant, and $m_e$ the electron mass. In this work, we
assume pure hydrogen gas.

The emitted bremsstrahlung photons are upscattered by inverse Compton
scattering. The amplification of photon energy by single scattering is
$A \equiv 1+4\theta_e+16\theta_e^2$. The probability of single
scattering is $P = 1 - \exp(-\tau_{es})$ (see e.g., Dermer et al.
1991). Therefore the mean amplification factor by Compton scattering
would be
\begin{equation}
   \eta_0 = 1-P + PA = 1 + P(A-1) .
\end{equation}
The fraction $1-P$ of photons remain unscattered while the fraction $P$
of photons are upscattered to $A$ times their initial energy. This
prescription is valid only for a single scattering Comptonization.
Since most CDAF flow considered here has $\tau_{es} < 1$, we use this
amplification factor for Comptonized bremsstrahlung. Although Dermer et
al. (1991) provides a handy formula for $\eta$, applicable to diverse
regimes, it can give incorrect values for $A \gg 1$ and $\tau_{es}
\lesssim 1$, the main parameter regime of CDAF. Although $\tau_{es}$ is
a hard-to-define quantity in a complex flow, we simply adopt $\tau_{es}
= n_e \sigma_T r$ in this work for the purpose of estimating the
Comptonization.

When $PA \gg 1$, Comptonization becomes saturated, and all photons that
are not absorbed are upscattered to $3T_e$, that is $\eta \rightarrow
\eta_{sat} = 3\theta_e /x$ where $x \equiv h\nu /m_e c^2$ (Dermer et
al. 1991). Then the fully saturated Comptonized bremsstrahlung emission
is
\begin{eqnarray}
  \Lambda_{Cbr}^{sat} &=& \int_{x_{abs}}^{3\theta_e} \eta_{sat}
  \epsilon_{br}(x)dx + \int_{3\theta_e}^\infty \epsilon_{br}(x)dx \\
  &=& \Lambda_{br} \left[ \int_3^\infty e^{-t} dt
  + \int_{x_{abs}/\theta_e}^3 3t^{-1}e^{-t}dt \right]
\end{eqnarray}
where $\epsilon_{br}(x)$ is the bremsstrahlung spectrum
\begin{equation}
  \epsilon_{br} (x) = \Lambda_{br}\exp\left(-\frac{x}{\theta_e}\right)
  dx
\end{equation}
and $x_{abs}$ is the absorption frequency $\nu_{abs}$ in unit of
$m_ec^2/h$. Since $\Lambda_{br}$ is smaller by the factor
$\exp(x_{abs}/\theta_e)$ in presence of absorption, we take the maximum
amplification factor for bremsstrahlung as
\begin{equation}
  \eta_{br}^{sat} = e^{x_{abs}/\theta_e} \times \left\{ e^{-3} + 3 \left[ E_1
  \left(\frac{x_{abs}}{\theta_e}\right) - E_1(3) \right] \right\},
\end{equation}
where $E_n \equiv \int_1^\infty t^{-n}\exp(-xt)dt$ is the exponential
integral. The final Comptonized bremsstrahlung emission is then
\begin{equation}\label{eq:LambCbr}
  \Lambda_{Cbr} = \min(\eta_0,\eta_{br}^{sat}) \Lambda_{br}.
\end{equation}

The energy-weighted mean photon energy for unsaturated Comptonized
bremsstrahlung is (see e.g. PO2)
\begin{equation}
  \frac{4 k T_X}{m_e c^2} = \frac{\int_0^\infty \eta_{br}^2 x
  \epsilon_{br}(x) dx}
  {\int_0^\infty \eta_{br} \epsilon_{br}(x) dx} = \eta_{br} \theta_e,
\end{equation}
and that for saturated Comptonized bremsstrahlung is simply
\begin{equation}
  T_X = T_e,
\end{equation}
since the spectrum approaches Wien spectrum which has $T_X = T_e$.
Therefore, the radiation temperature of locally radiated bremsstrahlung
emission is
\begin{equation}\label{eq:thetaX_local}
  T_X^{Cbr} = \frac{1}{4} T_e \min(\eta_{br},4).
\end{equation}

The frequency $\nu_{abs}$ is chosen to be the frequency at which the
free-free absorption optical depth is equal to 1,
\begin{equation}\label{eq:tauff}
  r a_{f\!f}(\nu_{abs}) = 1,
\end{equation}
where $a_{f\!f}$ is the absorption coefficient given by Dermer et al.
(1991)
\begin{equation}\label{eq:aff}
   a_{f\!f}(x) = \sqrt{8\pi}
             \frac{\alpha_f^2 \sigma_T r_e^3}{x^2 \theta_e^{3/2}
             \left[ 1 + (8/\pi)^{1/2}\theta_e^{3/2}\right]}
             n_i^2 \bar{g}
\end{equation}
with
\begin{eqnarray}
   \bar{g} & = & (1+2\theta_e+2\theta_e^2)
                 \ln \left[
                 \frac{4\eta_E(1+3.42\theta_e)\theta_e}{x} \right] \\
           & & + (\frac{3\sqrt{2}}{5}+2\theta_e)\theta_e
                 \ln \left[
                 \frac{4\eta_E(11.2+10.4\theta_e^2)\theta_e}{x} \right],
                 \nonumber
\end{eqnarray}
$\alpha_f$ is the fine structure constant, and $r_e = e^2/m_ec^2$ the
electron radius (Svensson 1984). This expression is valid for
$x\ll\theta_e$ and we used $(\pi/2)^{1/2}\theta_e^{1/2}[1 +
(8/\pi)^{1/2}\theta_e^{3/2}]$ to approximate
$\exp(1/\theta_e)\hbox{K}_2(1/\theta_e)$.

Flow at a given position is heated (or cooled) by the inverse
Comptonization off electrons. We use the heating rate (Levich \&
Sunyaev 1971)
\begin{equation}\label{eq:GammaC}
  \Gamma_C= 4 \sigma_T c [\theta_X(r) - \theta_e(r)] E_X(r) n_e(r,\vartheta) ,
\end{equation}
where $E_X(r)$ is the radiation energy density from equation
(\ref{eq:EXr}) and $\theta_X \equiv kT_X/m_e c^2$ the radiation
temperature from equation (\ref{eq:TXr}).

\subsection{Equilibrium Temperature and Overheating}

In original CDAF, the temperature of the gas is determined by the
balance between the viscous heating plus the convective energy
transport versus the radiative cooling. However, if the radiative
heating is dominant in some region of the flow (the condition under
which this assumption is valid is discussed in \S3.4.), the temperature
in that region will change to reach a new equilibrium. The new
equilibrium temperature of the flow will be determined by the balance
between radiative heating and radiative cooling.

In CDAF considered in this work, Compton heating versus Comptonized
bremsstrahlung cooling are the main radiative processes. The thermal
equilibrium temperature $T_{eq}$, then, satisfies
\begin{equation}\label{eq:Teq}
   \Gamma_C (T_{eq}) = \Lambda_{Cbr} (T_{eq})
\end{equation}
at given position $(r,\vartheta)$. From equations (\ref{eq:LambCbr})
and (\ref{eq:GammaC}), $\theta_{eq} \equiv kT_{eq}/m_e c^2$ is
determined by
\begin{equation}\label{eq:th_eq}
  4 cE_X(r) [\theta_X(r) - \theta_{eq}] = \alpha_f m_e c^3 n_e(r,\vartheta)
  \eta_{Cbr}(n_e,r,\theta_{eq}) [F_{ei}(\theta_{eq})+F_{ee}(\theta_{eq})] .
\end{equation}
Since the number density $n_e(r,\vartheta)$ decrease as $\vartheta
\rightarrow 0$ (toward the pole), the derived electron temperature
increases toward the pole as long as $\theta_X > \theta_e$.

For smaller enough $\vartheta$, the thermal equilibrium temperature can
be higher than the virial temperature,
\begin{equation}\label{eq:overheat}
  T_{eq} (r,\vartheta) > T_{vir} (r).
\end{equation}
The virial temperature is defined as $(5/2)kT_{vir} = m_p GM/r$, where
$m_p$ is the proton mass. Once electrons are heated above the virial
temperature, and therefore above the ion temperature, electrons are
likely (via collisions and instabilities) to heat ions to above the
virial temperature, thereby, creating winds, especially because the
dynamical time of CDAF is much longer than the free-fall flow.
Therefore, we adopt equation (\ref{eq:overheat}) as the condition for
overheating and producing a wind.

What equation (\ref{eq:overheat}) means is that some region of CDAF
will be radiatively heated from the background temperature \(T_e\) to
\(T_{eq}\) to achieve the thermal equilibiurm. However, as the
temperature of the flow approaches \(T_{eq}\), it will become unbound
when the condition (\ref{eq:overheat}) is met, and thereby producing
the outflow.

There is also a trivial, yet additional constraint for overheating: the
radiation temperature must be higher than the virial temperature
\begin{equation}\label{eq:TXTvir}
  T_X(r) > T_{vir}(r),
\end{equation}
otherwise the radiation field will cool the flow.

\subsection{Outflow}

Outflows will extend from the polar axis to the angle $\vartheta_c$ at
which the equilibrium temperature is equal to the virial temperature,
\begin{equation}\label{eq:thetac}
  T_{eq} (r,\vartheta_c) = T_{vir}(r).
\end{equation}
Regions with $\vartheta < \vartheta_c$ are overheated to above the
virial temperature. Hence, the shape of the outflow is determined by
the angle $\vartheta_c (r)$ as a function of $r$.

One example of the outflow is shown in Figure 3 for the mass accretion
rate of $\dot m = 10^{-3}$, the electron temperature of inner region
$T_1 = 10^{11} \K$. The {\sl solid curve} shows $\vartheta_c(r)$, and
the outer {\sl dotted circle} is the outer boundary of the flow. The
outflow starts at $r \simeq 40$ and the opening angle $\vartheta_c$
reaches maximum of $17^\circ$ around $r \simeq 700$ and ends with
$8.3^\circ$ at the outer boundary. Figure 4 shows the same for $\dot m
= 2.0 \times 10^{-3}$ and $T_1 = 3.0 \times 10^{10} \K$ with
$\vartheta_c = 6.6^\circ$ at the boundary. The inner isothermal region
and the immediate surroundings have $T_X < T_{vir}$, and the
overheating does not occur. However, at larger radius Compton heating
is strong enough to overheat the polar region of the gas and produce
outflow. The fraction of the sphere covered by the outflowing gas is
$2\pi\vartheta^2 / 4\pi \sim 0.01$ for $\vartheta=8.3^\circ$.

We have considered combinations of $\dot m$ and $T_1$. Outflow
solutions exist within a limited range of these parameters. Too high
$\dot m$ makes the outer boundary so small that the whole region is
isothermal, i.e., $T_X = T_e$, and the flow is not heated. The outer
boundary is smaller for higher $\dot m$ because radiative luminosity is
proportional to $\dot{m}^2$ while CDAF flows have a fixed (assumed)
radiation efficiency, which makes the total luminosity roughly
proportional to $\dot m$. The only way to reconcile this is $r_{out}$
being smaller for higher $\dot m$ (BNQ). In a too low $\dot m$ flow,
Comptonization is not strong enough to keep the radiation temperature
high. Lower energy photons from larger radii dilute the radiation field
and flows are not heated. In Figure 5, we show region of space in
($\dot m$,$T_1$) where outflow solutions are successfully produced.
{\sl Circles} denote ($\dot m$,$T_1$) for which outflows are found. The
size of circle represents the opening angle of the outflow at the outer
boundary, the majority being $\lesssim 10^\circ$. {\sl Crosses} denote
($\dot m$,$T_1$) values for which outflows do not exist.

As expected, flows with high electron temperature are more likely to
develop outflows. Outflows are not expected for $T_1 < m_ec^2/k \simeq
6\times 10^9 \K$: Comptonization is less efficient when $kT_e/m_e c^2 <
1$. Since the adopted radiation efficiency is $\epsilon_c \eta_c \simeq
10^{-2}$, we expect CDAFs with outflows to exist in the luminosity
range $8\times10^{-7} \lesssim L/L_E \lesssim 4\times10^{-5}$ for $T_1
= 10^{11}\K$ and $4\times10^{-5} \lesssim L/L_E \lesssim 10^{-4}$ for
$T_1 = 10^{10}\K$. These results do not depend on the mass of the black
holes.

So far, we have assumed that the total radiation efficiency $\epsilon_c
\eta_c$ is constant. In fact, current theoretical understanding of the
CDAF solution is not yet secure on this point. We now relax that
assumption and study an alternative case when $\epsilon_c \eta_c$ is
assumed to be proportional to $\dot m$. We adopt simply $\epsilon_c
\eta_c = 10^{-2} (\dot m/10^{-2}) $ so as to make $\epsilon_c \eta_c =
10^{-2}$ for $\dot m = 10^{-2}$. Now for small $\dot m$ the total
luminosity determined from equation (\ref{eq:LvsMdot}) and the total
bremsstrahlung emission from equation (\ref{eq:Lr}) are both
proportional to ${\dot m}^2$, and the outer boundary is now at a
roughly constant radius (see {\sl dotted lines} in Fig. 2). At high
$\dot m$, Comptonization adds additional power of $\dot m$ to the total
emission, and $r_{out}$ depends on $\dot m$ (see {\sl dotted lines} in
Fig. 2). The outflow solutions for this choice of radiation efficiency
are shown in Figure 6. Qualitatively, they are not much different from
the solutions with constant $\epsilon_c \eta_c$. Nonetheless, the flow
is much less extended due to lower radiation efficiency for lower $\dot
m$ and the shape of the outflow funnel in this case is almost straight.
The whole flow structure is quite self-similar except for the magnitude
of the opening angle. Most of the solutions still have the opening
angle at the outer boundary $\lesssim 20^\circ$.

\subsection{Convection, Heating, and Cooling Timescales}

In the original self-similar CDAF, the viscous heating is balanced by
the convective energy transport. However, radiative heating can be
dominant in some region of the flow. The importance of each heating and
cooling process is measured by the corresponding timescale. The
radiative heating timescale is given by the ratio between the internal
energy of the electron, \(\varepsilon_e\), and the Compton heating
rate,
\begin{equation}\label{eq:t_H}
t_H \equiv \frac{\varepsilon_e}{\Gamma_C}.
\end{equation}
The radiative cooling timescale is similarly given by
\begin{equation}\label{eq:t_C}
t_C \equiv \frac{\varepsilon_e}{\Lambda_{Cbr}}.
\end{equation}
The timescale for convective energy transport at a certain
two-dimensional position is not trivial to evaluate  because the
two-dimensional velocity profile including the convective motion is
needed. Although two-dimensional numerical simulations (Stone, Pringle,
\& Begelman 1999; Igumenshchev, Abramowicz, \& Narayan 2000;
Igumenshchev \& Abramowicz 2000) suggest very limited flow motion near
the pole as in self-similar two-dimensional ADAF (Narayan \& Yi 1995),
the two-dimensional flow motion has yet to be expressed in simple
analytic form. Here, we adopt the height-averaged convective energy
flux given by NIA,
\begin{equation}\label{eq:F_c}
F_c = - \alpha_c \frac{c_s^2}{\Omega_K} \rho T \frac{ds}{dr},
\end{equation}
where \(\alpha_c\) is the convection coefficient analogous to the usual
Shakura \& Sunyaev \(\alpha\), \(c_s\) the isothermal sound speed,
\(\Omega_K\) the Keplerian angular velocity, and \(s\) the entropy of
the flow. In a self-similar solution that is marginally stable to
convection, \( \alpha_c = 3 \alpha \) is expected when \( \alpha
\lesssim 0.05 \) (NIA). The cooling rate per volume due to this
convective energy flux is then
\begin{equation}\label{eq:Q_cv}
Q_{cv} \equiv \frac{F_c}{r}.
\end{equation}
Corresponding timescale \(t_{cv}\) is now
\begin{equation}\label{eq:t_cv}
t_{cv} = \frac{\varepsilon_i}{Q_{cv}}.
\end{equation}
Since the actual convective motion near the pole is generally much
smaller than the average convective motion near the disk midplane as is
assumed in equation (\ref{eq:F_c}), the actual timescale for the
convective cooling could be much longer than \(t_{cv}\) evaluated by
equation (\ref{eq:t_cv}).

For given dimensionless mass accretion rate \(\dot m\) and the inner
electron temperature \(T_1\), small \(\alpha\) (of Shakura-Sunyaev), or
equally small \(\alpha_c\), means slower radial motion, and therefore,
slower convective motion. So we searched for the critical value of
\(\alpha\) below which \(t_{cv} > t_H \) is satisfied within the
overheated region. For higher electron temperature of \(T_1 = 10^{11}
\K\). the value of the critical \(\alpha\) ranges from \(2.0 \times
10^{-3}\) for \(\dot m = 2 \times 10^{-3}\) to \(1.2 \times 10^{-5}\)
for  \(\dot m = 8 \times 10^{-5}\), all at a constant radiation
efficiency of \(10^{-2}\). For lower electron temperature of \(T_1 = 3
\times 10^{10} \K\). the value of the critical \(\alpha\) ranges from
\(2.7 \times 10^{-2}\) for \(\dot m = 8 \times 10^{-3}\) to \(3.0
\times 10^{-4}\) for  \(\dot m = 8 \times 10^{-4}\). For even lower
electron temperature of \(T_1 = 10^{10} \K\) at which overheated region
is very limited in radius, the value of the critical \(\alpha\) ranges
from \(1.6 \times 10^{-2}\) for \(\dot m = 10^{-2}\) to \(2.8 \times
10^{-3}\) for \(\dot m = 4 \times 10^{-3}\). Expected values of
\(\alpha\) for which CDAF exists are in the range \(\alpha < 0.05 \)
(NIA). We, therefore, conclude that the Compton heated outflow is
expected to be produced in CDAF if the viscosity parameter \(\alpha\)
is small enough, exact value being dependent on the mass accretion
rate. However, as already mentioned, the real convective motion near
the pole is expected to be much smaller than the average convective
motion in the disk, the Compton heated outflow near the pole may
develop even for much larger value of \(\alpha\).

Another timescales we want to check are \(t_H\) versus \(t_C\) at the
temperature of the background flow \(T_e(r,\vartheta)\) given by
equation (\ref{eq:Te_r}). Only when radiative heating is greater than
cooling, i.e., \(t_H|_{T_e} < t_C|_{T_e}\), the flow is heated from
\(T_e\) to overheated temperature \(T_{eq}\) of equation
(\ref{eq:Teq}). The ratio \((t_H/t_C)_{T_e}\) can be expressed as
\begin{equation}\label{eq:tH_tC_Te}
  \frac{t_C}{t_H} = \frac{\Gamma_C(T_e)}{\Lambda_{Cbr}(T_e)} =
  \frac{\Gamma_C(T_{eq})}{\Lambda_{Cbr}(T_{eq})}
  \frac{\Gamma_C(T_e)}{\Gamma_C(T_{eq})}
  \frac{\Lambda_{Cbr}(T_{eq})}{\Lambda_{Cbr}(T_e)} =
  \frac{\Gamma_C(T_e)}{\Gamma_C(T_{eq})}
  \frac{\Lambda_{Cbr}(T_{eq})}{\Lambda_{Cbr}(T_e)},
\end{equation}
which is always less than 1 as long as \(T_{eq}>T_e\) because
\(\Gamma_C(T_e)\) is a monotonically decreasing function of \(T_e\) and
\(\Lambda_{Cbr}(T_e)\) a monotonically increasing function of \(T_e\).
The ratio \(\Gamma_C(T_{eq})/\Lambda_{Cbr}(T_{eq})=1\) by the
definition of \(T_{eq}\) in equation (\ref{eq:Teq}).

\subsection{Synchrotron Emission}

The synchrotron emission of hot electrons can produce copious soft
photons some of which are subsequently Compton upscattered. This will
lead to an increase in the total cooling rate as well as a decrease in
the radiation temperature of the emitted radiation. Increased cooling
rate will reduce the outer boundary of the CDAF whereas decreased
radiation temperature will reduce the Compton heating, or even cool the
flow if the radiation temperature falls below the electron temperature.

So we recalculated the condition for overheating in presence of
magnetic field for the flow parameters considered in Fig. 5. The
treatment of Comptonized synchrotron is described in Appendix B. We fix
the black hole mass to be $10^8 M_\odot$. We find that the synchrotron
emission can significantly affect the high temperature flow whereas the
low temperature flow is less affected. The overheating and subsequent
production of the outflow is still possible for flow with \(T_1 =
10^{10} \K\) as long as the magnetic field is less than 10\% of the
equipartition (the gas pressure being equal to the magnetic pressure)
field. The outer boundary of the flow and the opening angle of the
outflow are hardly changed if this condition is met. For \(T_1 = 3
\times 10^{10} \K\), we find that the magnetic field has to be smaller
than 3\% of the equipartition field to produce the outflow. For the
highest temperature \(T_1 = 10^{11} \K\) flow, the production of the
outflow is possible only when the magnetic field is less than 1\% of
the equipartition field. So we conclude that the Compton preheated
outflow is possible in the magnetic CDAFs as long as the magnetic field
is less than from 10\% to 1\% of the equipartition field.

\section{Summary and Discussion}

Hot accretion flows like ADAFs have a number of physical
characteristics that compliment the classic low-temperature disk flows.
In previous work, we have explored the consequences of the high
temperature and the two-dimensional density structure of these flows,
and have found that ADAFs may be able to produce radiatively driven
outflows. We subsequently have noted that self-similar CDAFs have even
more suitable properties for producing outflows as compared to ADAFs:
steeper poloidal density gradients and higher radiation efficiencies.
In this paper, we have studied the conditions for self-similar
two-dimensional CDAFs (NIA; QG) to develop radiatively heated polar
outflows.

1. We have found that CDAFs produce enough luminosity and photon energy
to drive polar outflows via Compton heating for a reasonable range of
mass accretion rate, or, equally, luminosity, as long as the magnetic
field is less than from 10\% to 1\% of the equipartition field. When
the electron temperature saturates around $10^{11}\K$ at the inner
region, polar outflows are possible for $8\times10^{-7} \lesssim L/L_E
\lesssim 4\times10^{-5}$ for radiation efficiency of $10^{-2}$, where
$L_E$ is the Eddington luminosity. The luminosity range for which
outflow exists is narrower for lower electron temperature flows and
disappears completely for electron temperature $\lesssim 6\times
10^9\K$

2. In most cases, outflows are well collimated along the rotation axis,
with an opening angle typically in the range $\lesssim 10^\circ$.

3. If we, instead of taking efficiency as constant, assume that it is
proportional to the mass accretion rate $\dot m$, the solutions are
qualitatively the same but are more self similar, i.e., opening angle
and outer boundary (in Schwarzschild units) depend less strongly on
$\dot m$.

4. Outflow is more probable for small viscosity parameter \(\alpha\).

The treatment in this work is not completely satisfactory in the sense
that the dynamics and the temperature profiles of CDAFs are not
self-consistently solved. However, it was not our intention to solve
fully three-dimensional, self-consistent gloabl CDAFs with proper
consideration for all gas, radiative, and magnetic processes, which
will be eventually needed to fully understand CDAFs. Rather, our goal
has been to show that even in the framework of simple self-similar
solutions, radiatively heated polar outflows appear as natural
consequences of the physical characteristics of CDAFs.

\acknowledgments

This work is the result of research activities (Astrophysical Research
Center for the Structure and Evolution of the Cosmos) supported by
Korea Science \& Engineering Foundation.

\appendix

\section{Radiation Field Inside An Optically Thin Spherical Shell}

Let's consider an optically thin, uniform radiating shell with radius
$r'$ and the thickness $\Delta r'$. The specific intensity at radius
$r$ ($r<r'$) in the direction of $\vartheta$ (see Figure 7) is given by
integration of radiative transfer equation along that direction. If we
denote the emissivity (per unit volume per unit solid angle) of the
shell as $\epsilon_\nu/4\pi$, then the specific intensity is simply the
emissivity times the path length,
\begin{equation}
   I_\nu = \frac{\epsilon_\nu}{4\pi} \Delta s
         = \frac{\epsilon_\nu}{4\pi} \frac{\Delta r'}{\cos \phi},
\end{equation}
where $\phi$ is the angle between the ray and the normal of the shell.
From the law of sines, $\sin\phi = (r/R)\sin\vartheta$. The radiation
energy density, $E_\nu(r)$, is given by the integral of $I_\nu$ over
the solid angle $d\Omega$,
\begin{eqnarray}\label{eq:Enu}
   E_\nu (r) &=& \frac{1}{c} \int \frac{\epsilon_\nu}{4\pi} \frac{\Delta r'}{\cos\phi}
                 d\Omega \\
             &=& \frac{1}{c} \frac{\epsilon_\nu}{4\pi} \Delta r' \int
                 \frac{\sin\vartheta d\vartheta}
                 {\sqrt{1-\frac{r^2}{r'^2}\sin^2\vartheta}} d\varphi \\
             &=& \frac{1}{c} \epsilon_\nu \Delta r'
                 \frac{r'}{r}\ln\sqrt{\frac{r'+r}{r'-r}}.
\end{eqnarray}
Equation (\ref{eq:Enu}) formally diverges when $r \rightarrow r'$
whereas the correct value saturates. In the limit when $r'-r \lesssim
\Delta r'$, the effect of shell's curvature has to be incorporated in
$\Delta s$. However, since the radiation energy density is calculated
by the sum of contributions from discrete shells in $r$ (eq.
[\ref{eq:EXr}]), the divergence is automatically avoided.

\section{Comptonized Synchrotron Emission}

The angle-averaged synchrotron emission by relativistic Maxwellian
electrons is given by (Pacholczyk 1970)
\begin{equation}\label{eq:ep_syn}
   \epsilon_{syn}(\nu)d\nu = \frac{2\pi}{\sqrt{3}} \frac{e^2}{c}
                    \frac{n_e\nu}{\theta_e^2} I'(x_M) d\nu
\end{equation}
where $x_M \equiv 2\nu/(3\nu_0\theta_e^2)$, $\nu_0 \equiv eB/(2\pi m_e
c)$, and
\begin{equation}\label{eq:Iprime}
   I'(x_M) = \frac{4.0505}{x_M^{1/6}}
             \left(1 + \frac{0.40}{x_M^{1/4}}
                   + \frac{0.5316}{x_M^{1/2}}\right)
             \exp(-1.8899x_M^{1/3})
\end{equation}
is a fitting formula (Mahadevan, Narayan, \& Yi 1996). When absorption
is not important, the cooling rate due to the optically thin
synchrotron emission is obtained by integrating the equation
(\ref{eq:ep_syn})
\begin{equation}\label{eq:lambsyn0}
   \Lambda_{syn}^0 = 213.6 \frac{e^2}{c} n_e \nu_0^2 \theta_e^2.
\end{equation}

However, a large fraction of the low energy synchrotron photons are
generally absorbed by synchrotron self-absorption. The synchrotron
emission in the presence of absorption can be approximated as
\begin{equation}\label{eq:lambsyn}
   \Lambda_{syn} = f_{syn} \Lambda_{syn}^0 ,
\end{equation}
where
\begin{equation}\label{eq:fsyn}
   f_{syn} \equiv \int_{x_M^{abs}}^\infty x_M I'(x_M)dx_M
           \Big/ \int_0^\infty x_M I'(x_M)dx_M
\end{equation}
is the fraction of synchrotron emission above the synchrotron
self-absorption frequency $\nu_{abs}$ which satisfies
\begin{equation}
   \tau_{syn} = \frac{1}{4\sqrt{3}}
                \frac{e^2 c n_e(r) r}{\nu_{abs} kT_e \theta_e^2}
                I'(\nu_{abs}) = 1.
\end{equation}
Thus, we only consider the optically thin part of synchrotron emission
as the cooling function for the gas and as a contribution to the
preheating radiation field.

Locally emitted synchrotron photons are upscattered by inverse
Comptonization off hot electrons. Here, we adopt a simple estimate of
Comptonized synchrotron, which is reasonable in physical conditions
considered in this paper,
\begin{equation}\label{eq:lambcs}
   \Lambda_{CS} = \eta_{syn} f_{syn} \Lambda_{syn}^0,
\end{equation}
where
\begin{equation}\label{eq:etasyn}
   \eta_{syn} = \min ( \eta_0 , \eta_{syn}^{sat} )
\end{equation}
is the Comptonized energy enhancement factor (see e.g., Dermer et al.
1991) for synchrotron emission. The enhancement factor for fully
saturated Comptonized synchrotron in the presence of absorption is
\begin{equation}\label{eq:etasynmax}
   \eta_{syn}^{sat} = \frac{\int_{x_M^{abs}}^{\infty}
                      \frac{3kT_e}{h\nu} x_M I'(x_M) dx_M}
                      {\int_{x_M^{abs}}^{\infty} x_M I'(x_M) dx_M},
\end{equation} (see \S3.1 for bremsstrahlung case). The synchrotron absorption
frequency is compared with that for free-free absorption, and the
larger of the two is chosen.

The radiation temperature of locally emitted Comptonized synchrotron is
similarly
\begin{equation}\label{eq:TXCS}
   T_X^{CS} = \min (\eta_0 \frac{3}{8}\frac{h\nu_0}{k}\theta_e^2
                  < \! x_M \! >_{\epsilon_\nu}, T_e) ,
\end{equation}
where
\begin{equation}\label{eq:xMavg}
   < \! x_M \! >_{\epsilon_\nu} \equiv  \frac{\int_{x_M^{abs}}^\infty
                      x_M^2 I'(x_M) dx_M }
                      {\int_{x_M^{abs}}^\infty x_M I'(x_M) dx_M }
\end{equation}
is the energy-weighted mean photon energy in $x_M$ unit. This treatment
of Comptonized synchrotron radiation is almost the same as that adopted
in Park \& Ostriker (2001), except that the energy enhancement factor
and the treatment of saturated Comptonization are slightly improved.

In the presence of both bremsstrahlung and synchrotron emission,
$\Lambda$ is replaced by $\Lambda_{Cbr} + \Lambda_{CS}$ in equation
(\ref{eq:EXr}) and $T_X \Lambda$ by $T_X^{Cbr}\Lambda_{Cbr} +
T_X^{CS}\Lambda_{CS}$ in equation (\ref{eq:TXr}).

\clearpage

\begin{figure}
\plotone{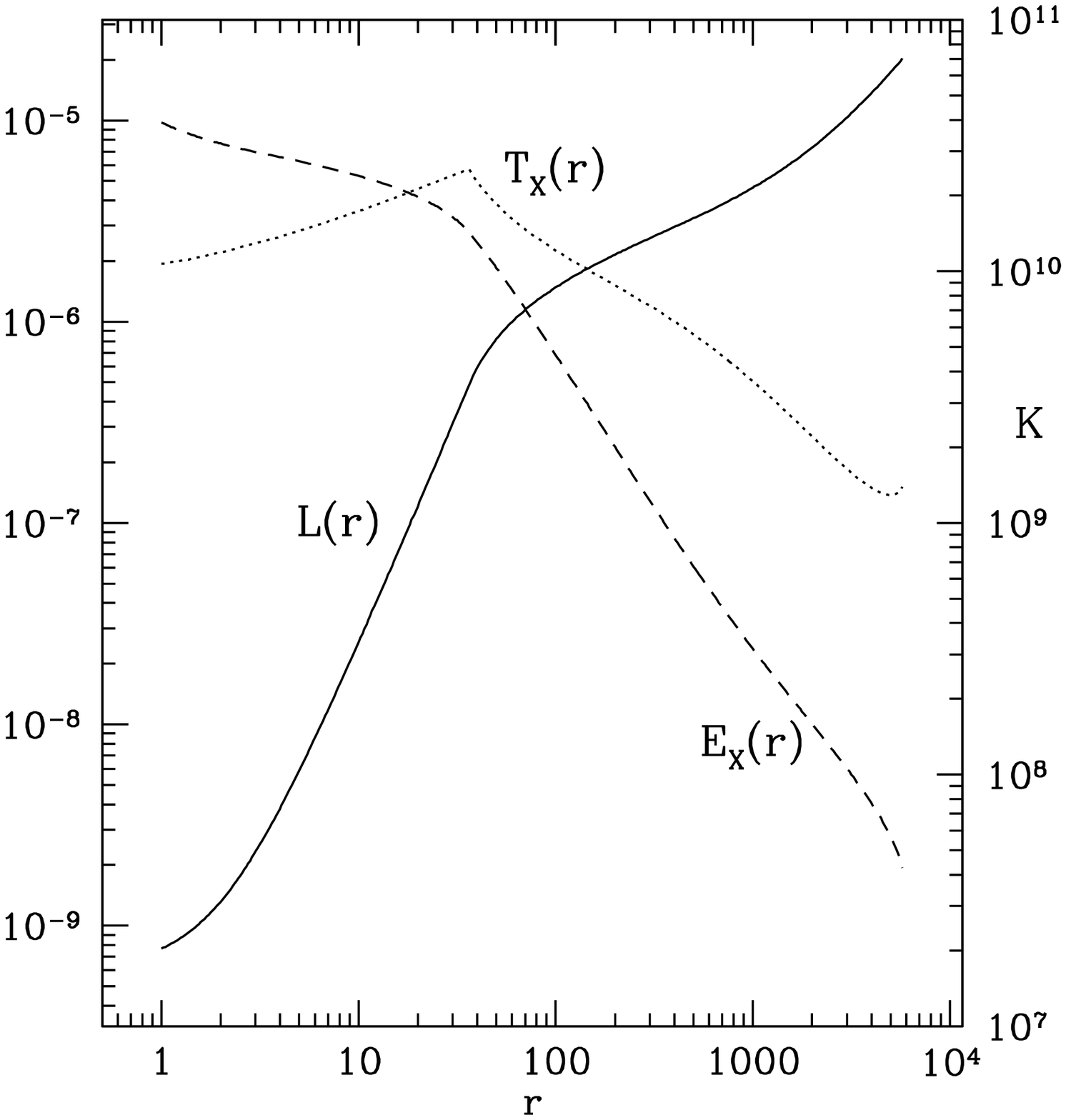} \caption{Typical log profiles of luminosity $L(r)$ (in
units of $L_E$, left ordinate), of the radiation energy density
$E_X(r)$ (in arbitrary units, left ordinate), and of the radiation
temperature $T_X(r)$ (in K, right ordinate).}
\end{figure}

\begin{figure}
\plotone{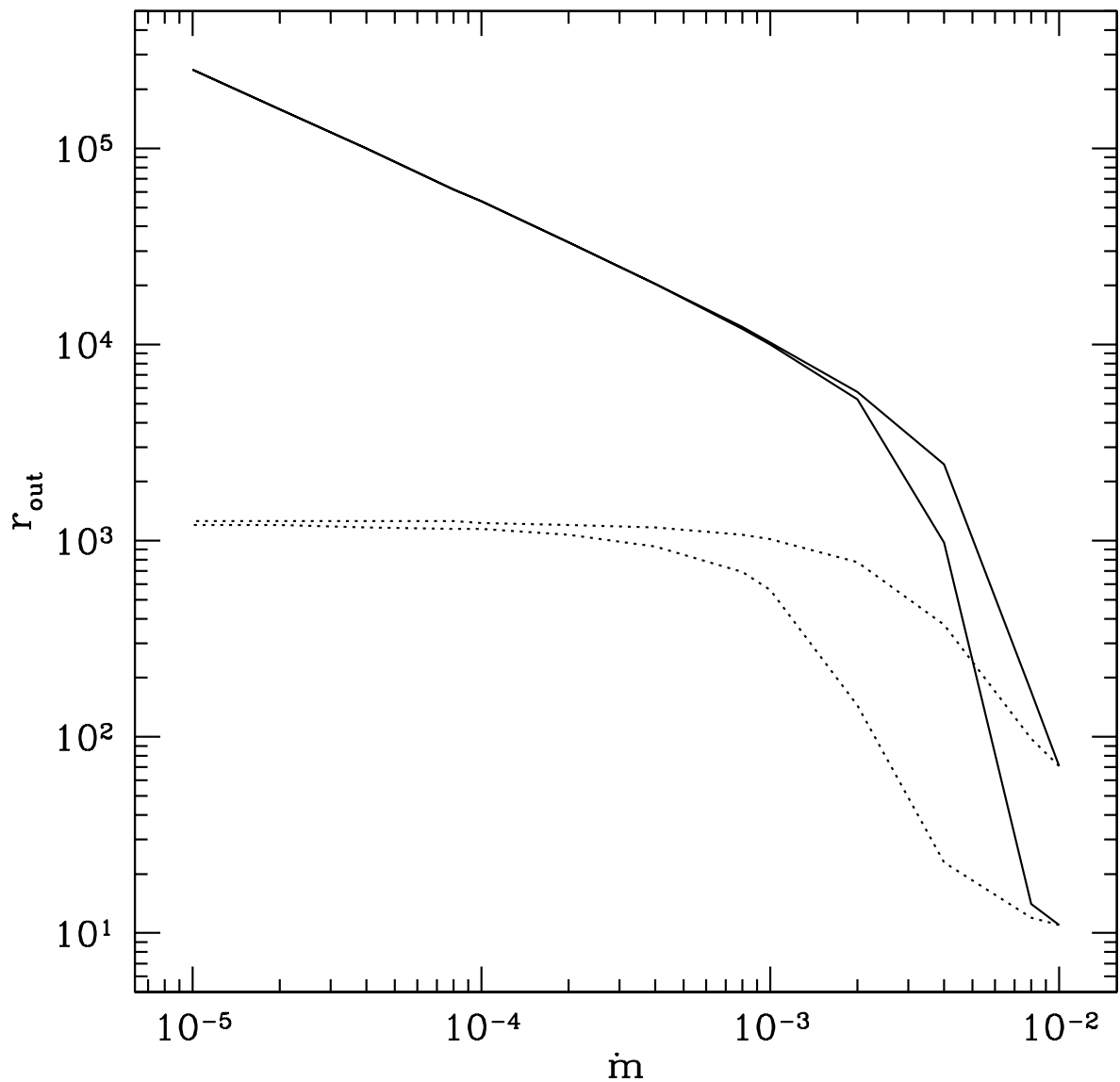} \caption{The outer boundary radius $r_{out}$ as a
function of $\dot m$. {\sl Upper solid line} is for $T_1 = 3 \times
10^{10} \K$ and {\sl lower one} for $T_1 = 10^{11}\K$, both for
constant radiation efficiency $\epsilon_c \eta_c = 0.01$. {\sl Dotted
lines} represent the same $r_{out}$ for varying radiation efficiency
$\epsilon_c \eta_c = 10^{-2} (\dot m/10^{-2})$: {\sl upper one} for
$T_1 = 3 \times 10^{10} \K$ and {\sl lower one} for $T_1 = 10^{11}\K$.}
\end{figure}

\begin{figure}
\plotone{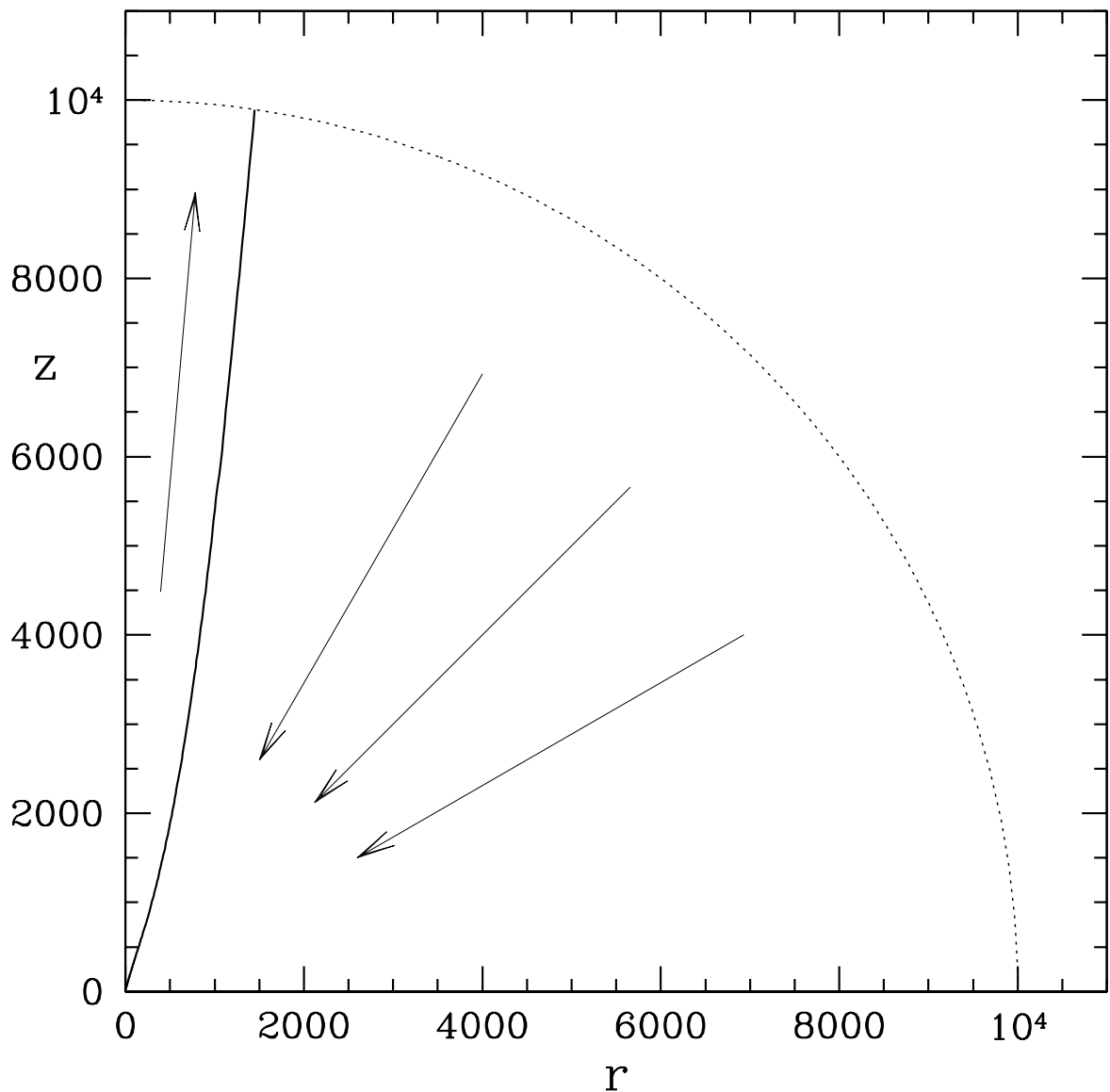} \caption{The region inside the {\sl solid curve}
toward the pole is overheated above the virial temperature due to
Compton heating. Outer {\sl dotted circle} shows the outer boundary of
CDAF for given total luminosity. This figure is for $\dot{m}=10^{-3}$
and $T_1=10^{11}\K$. The opening angle of the outflow at the outer
boundary is only $8^\circ$.}
\end{figure}

\begin{figure}
\plotone{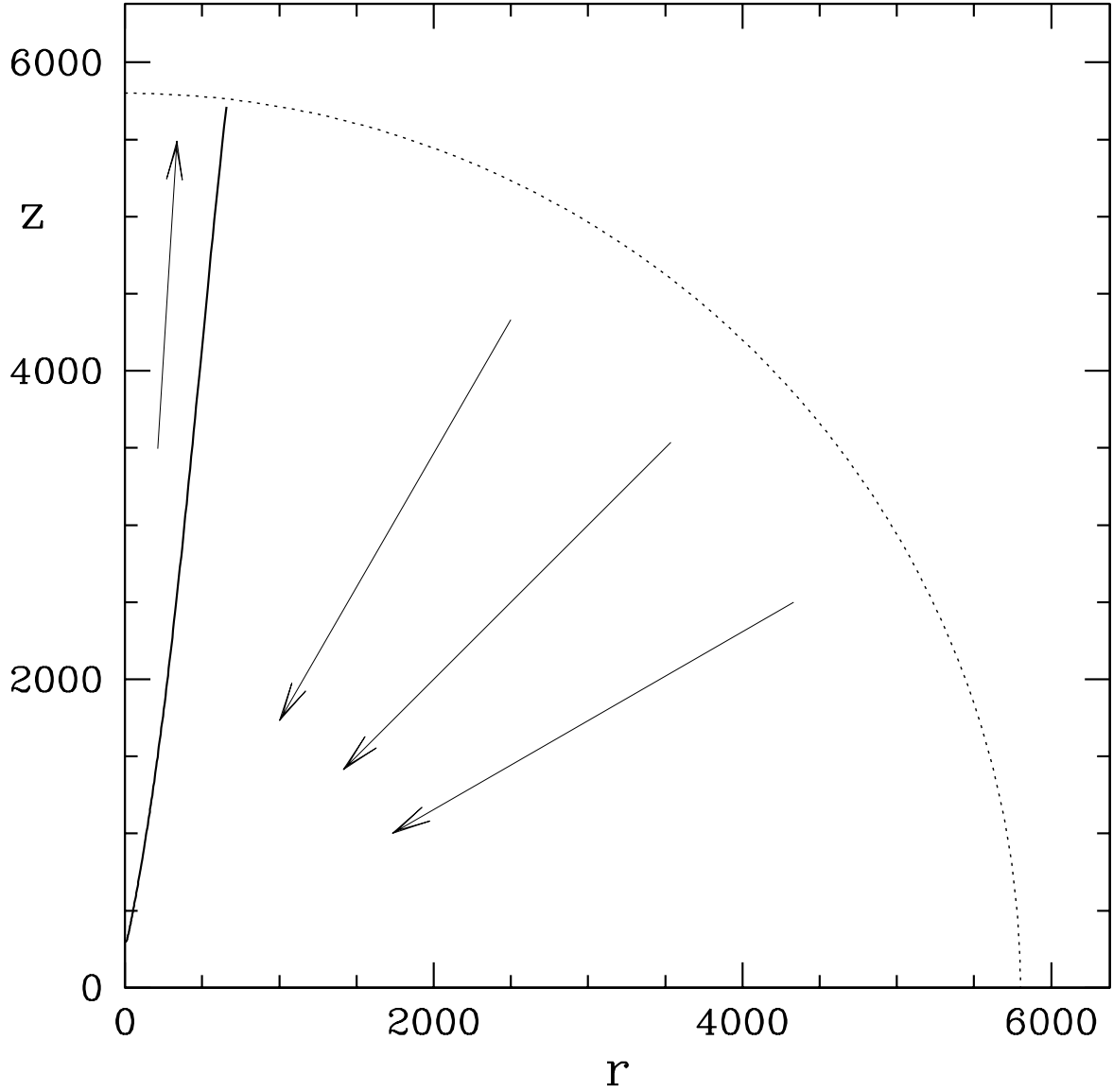} \caption{Same as Figure 3, but for
$\dot{m}=2\times10^{-3}$ and $T_1= 3\times10^{10}\K$. The opening angle
at the outer boundary is $8^\circ$.}
\end{figure}

\begin{figure}
\plotone{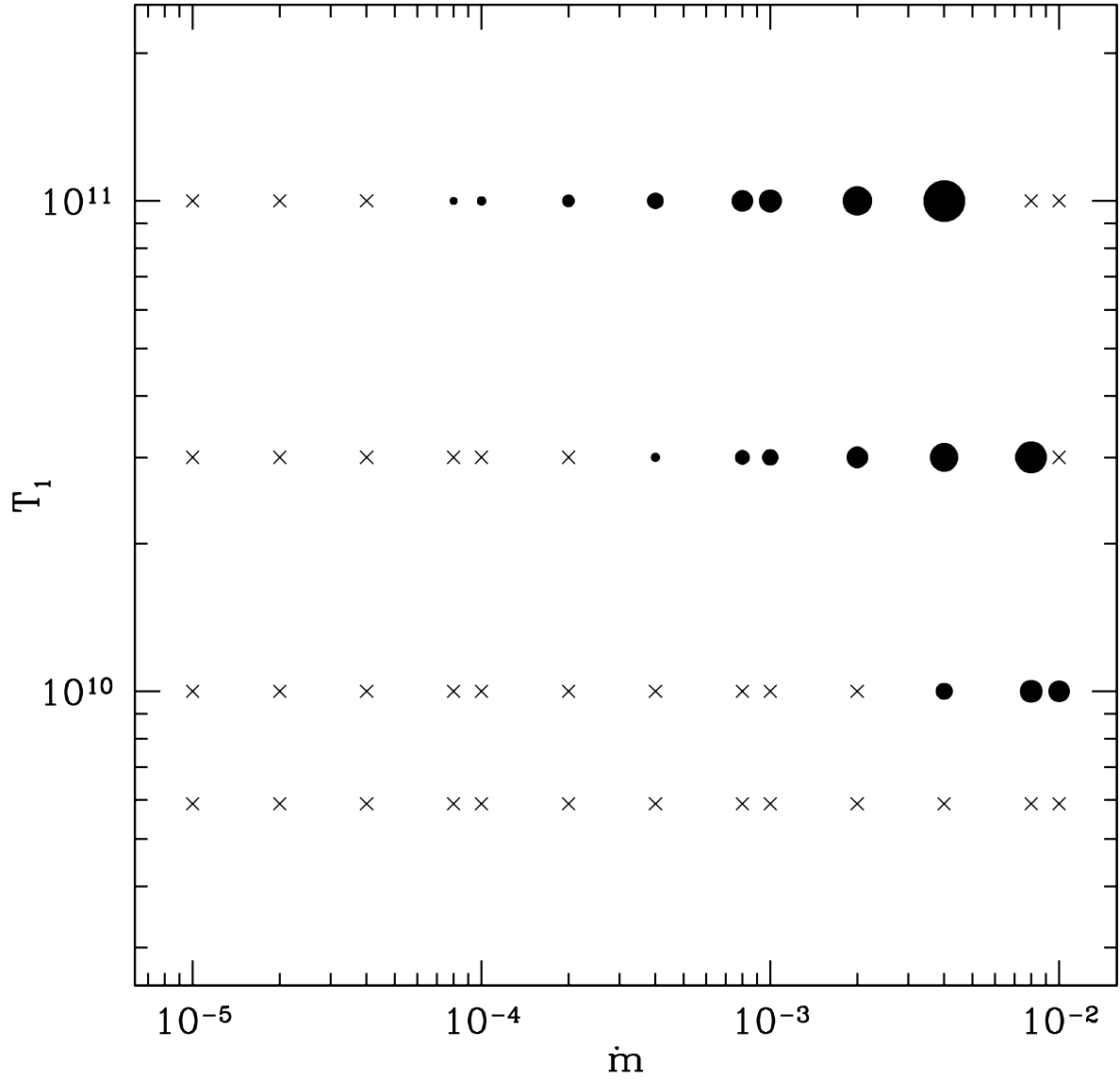} \caption{ The parameter space of ($\dot m$, $T_1$)
under constant radiation efficiency for which outflow exists is denoted
as {\sl circles}. The size of circle represents the opening angle of
the outflow at the outer boundary. No outflow is found at ($\dot m$,
$T_1$) denoted by {\sl crosses}. }
\end{figure}

\begin{figure}
\plotone{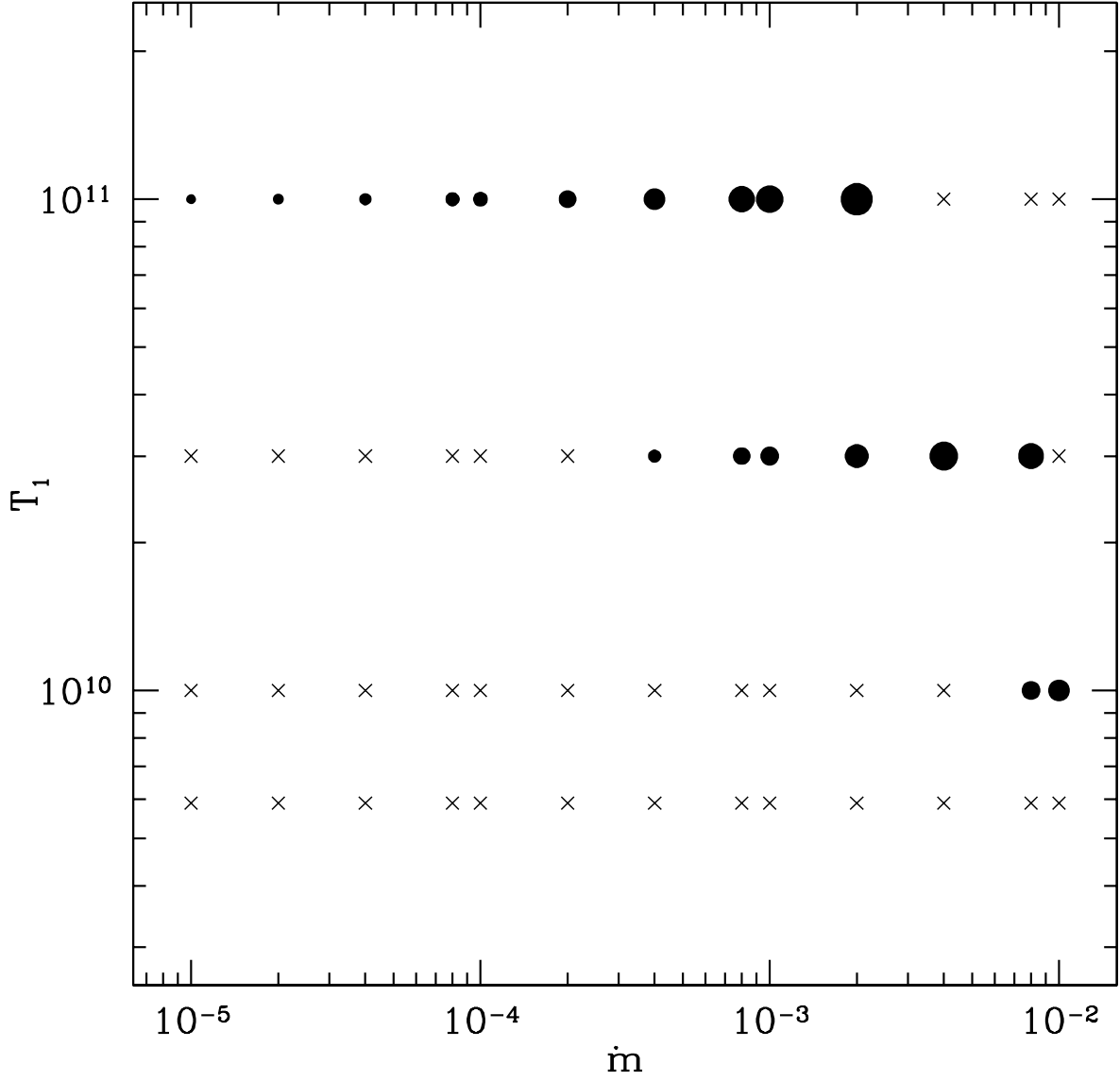} \caption{ Same as Figure 5, but now the radiation
efficiency is proportional to the mass accretion rate. }
\end{figure}

\begin{figure}
\plotone{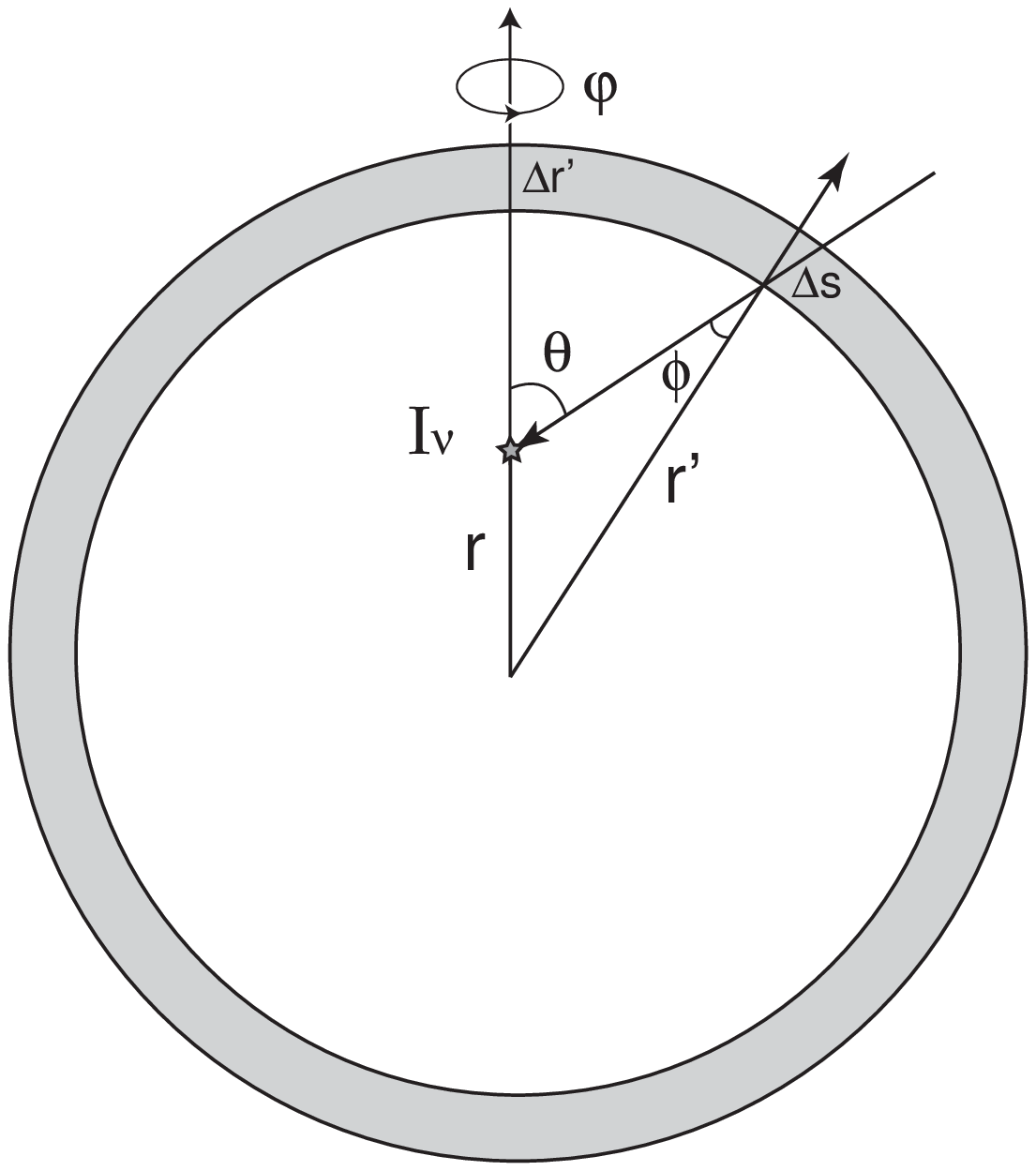} \caption{The diagram for calculation of specific
intensity, $I_\nu$, inside a radiating spherical shell.}
\end{figure}


\begin{references}

\reference{} Abramowicz, M., Chen, X., Kato, S., Lasota, J.-P.,
             \& Regev, O. 1995, \apjl, 438, L37
\reference{} Ball, G.~H., Narayan, R., \& Quataert, E.\ 2001,
             \apj, 552, 221 (BNQ)
\reference{} Begelman, M. C., \& Meier, D. L. 1982, \apj, 253, 873
\reference{} Blandford, R. D., \& Begelman, M. C. 1999, \mnras, 303, L1
\reference{} Ciotti, L., \& Ostriker, J. P. 1997, \apjl, 487, L105
\reference{} Ciotti, L., \& Ostriker, J. P. 2001, \apj, 551, 131
\reference{} Cowie, L. L., Ostriker, J. P., and Stark, A. A. 1978,
             \apj, 226, 1041
\reference{} Dermer, C. D., Liang, E. P., \& Canfield, E. 1991, \apj, 369, 410
\reference{} Ichimaru, S. 1997, \apj, 214, 840
\reference{} Igumenshchev, I. V., \& Abramowicz, M. A. 1999, \mnras, 303, 309
\reference{} Igumenshchev, I. V., \& Abramowicz, M. A. 2000, \apjs, 130, 463
\reference{} Igumenshchev, I. V., Abramowicz, M. A., \& Narayan, R. 2000,
             \apjl, 537, L27
\reference{} Igumenshchev, I. V., Chen, X., \& Abramowicz, M. A. 1996,
             \mnras, 278, 236
\reference{} Levich, E.~V.~\& Syunyaev, R.~A.\ 1971, Soviet Astronomy, 15, 363
\reference{} Mahadevan, R., Narayan, R., \& Yi, I. 1996, \apj, 465, 327
\reference{} Narayan, R., Barret, D., \& McClintock, J.~E.\ 1997, \apj, 482, 448
\reference{} Narayan, R., Igumenshchev, I. V., \& Abramowicz, M. A., 2000,
             \apj, 539, 798 (NIA)
\reference{} Narayan, R., Mahadevan, R., \& Quataert 1999,
             in The Theory of Black Hole Accretion Discs,
             eds. M. A. Abramowicz, G. Bjornsson, and J. E. Pringle
             (Cambridge: Cambridge U. Press), 148 (astro-ph/9803141)
\reference{} Narayan, R., \& Yi, I. 1994, \apj, 428, L13
\reference{} Narayan, R., \& Yi, I. 1995a, \apj, 444, 231
\reference{} Narayan, R., \& Yi, I. 1995b, \apj, 452, 710
\reference{} Nobili, L., Turolla, R., \& Zampieri, L. 1991, \apj, 383, 250
\reference{} Ostriker, J. P., McCray, R., Weaver, R., \& Yahil, A.
             1976, \apjl, 208, L61
\reference{} Pacholczyk, A. G. 1970, Radio Astrophysics
             (San Francisco: Freeman)
\reference{} Park, M.-G. 1990a, \apj, 354, 64
\reference{} Park, M.-G. 1990b, \apj, 354, 83
\reference{} Park, M.-G. \& Ostriker, J. P. 1999, \apj, 527, 247
\reference{} Park, M.-G. \& Ostriker, J. P. 2001, \apj, 549, 100
\reference{} Quataert, E., \& Gruzinov, A. 2000, \apj, 539, 809 (QG)
\reference{} Rees, M. J., Begelman, M. C., Blanford, R. D.,
             \& Phinney, E. S. 1982, \nat, 295, 17
\reference{} Shakura, N. I., \& Sunyaev, R. A. 1973, \aap, 24, 337
\reference{} Stepney, S., \& Guilbert, P.~W. 1983, \mnras, 204, 1269
\reference{} Stone, J. M., Pringle, J. E., \& Begelman, M. C. 1999,
             \mnras, 310, 1002
\reference{} Svensson, R. 1982, \apj, 258, 335
\reference{} Svensson, R. 1984, \mnras, 209, 175
\reference{} Wandel, A., Yahil, A., \& Milgrom, M. 1984, \apj, 282, 53
\reference{} Xu, G., \& Chen, X. 1997, \apjl, 489, L29
\reference{} Zampieri, L., Miller, J. C., \& Turolla, R. 1996, \mnras, 281, 1183

\end{references}
\end{document}